\documentclass[journal]{vgtc}                     

\graphicspath{{figures/}{pictures/}{images/}{./}} 

\usepackage{times}                     

\usepackage{tabu}                      
\usepackage{booktabs}                  
\usepackage{lipsum}                    
\usepackage{mwe}                       
\usepackage{xcolor}
\usepackage[percent]{overpic}
\usepackage{graphicx}
\usepackage{subcaption}
\usepackage{svg}

\onlineid{0}

\vgtccategory{Research}

\vgtcpapertype{evaluation}

\title{Latency Effects on Multi-Dimensional QoE in Networked VR Whiteboards}

\author{%
  \authororcid{Jiarun Song}{0000-0001-6718-4201},
  Yongkang Hou, and
  Fuzheng Yang
}

\authorfooter{
  \item
  	Jiarun Song is with School of Telecommunications Engineering, Xidian University.
  	E-mail: jrsong@xidian.edu.cn (Corresponding author)
  \item
  	Yongkang Hou is with School of Telecommunications Engineering, Xidian University.
  	E-mail: waterhou1221@gmail.com.

  \item Fuzheng Yang is with School of Telecommunications Engineering, Xidian University.
  	E-mail: fzhyang@mail.xidian.edu.cn.
}

\abstract{%
  Networked virtual reality (NVR) whiteboards are increasingly important for enabling geographically dispersed users to engage in real-time idea sharing, collaborative design, and discussion. However, latency caused by network limitations, rendering delays, or synchronization issues can significantly degrade the Quality of Experience (QoE) in whiteboard collaboration. To systematically investigate the impact of latency, this study classified QoE into pragmatic and hedonic aspects, each comprising multiple sub-dimensions. Controlled experiments were conducted to identify the sub-dimensions most affected by latency, which were then adopted as the primary QoE indicators, with the aim of uncovering the processes and mechanisms through which latency shapes QoE. Building on this, we further examined how these impacts vary across different collaboration modes, namely sequential collaboration (SC) for structured design workflows and free collaboration (FC) for open discussion. We also compared two VR whiteboard types, one with avatars (VR+) and the other without avatars (VR), and included a traditional PC-based whiteboard as a baseline. This multi-dimensional design enables a comprehensive evaluation of latency’s impact on QoE across collaboration modes and platforms, providing practical guidance for optimizing NVR whiteboard systems under real-world network and system constraints.
}


\teaser{
  \centering
  \begin{overpic}[width=\linewidth]{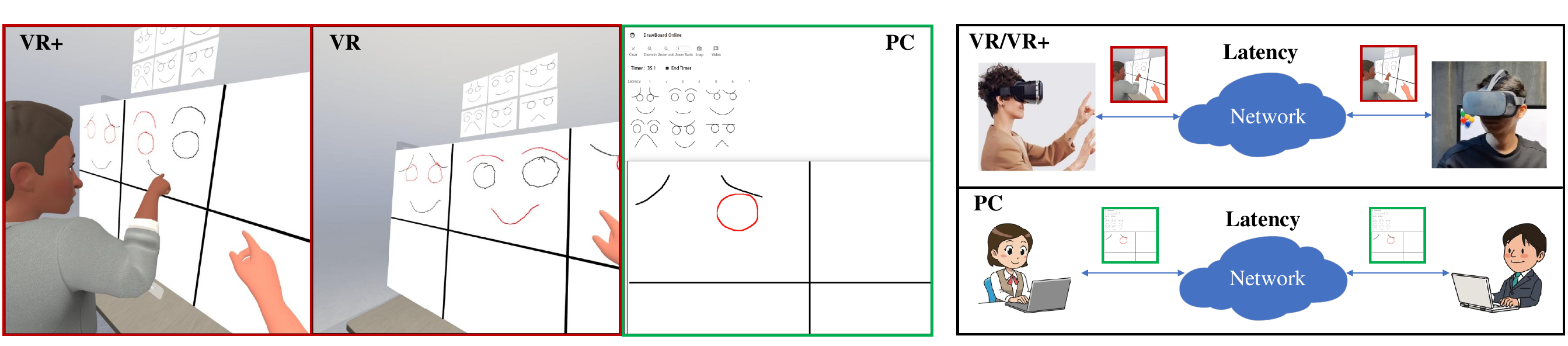}
    \put(9,-2){\small{(a)}}      
    \put(29,-2){\small{(b)}}
    \put(49,-2){\small{(c)}}
    \put(80,-2){\small{(d)}}
  \end{overpic}
   \vspace{0.5pt}
  \caption{Collaborative whiteboard platforms and experimental setup used in the study.
  (a) Networked VR whiteboard with embodied avatars (VR+).
  (b) Networked VR whiteboard without avatars (VR).
  (c) PC-based collaborative whiteboard (PC).
  (d) VR/VR+ and PC collaboration scenarios under network latency.}
  \label{fig:Fig1}
}

\graphicspath{{figs/}{figures/}{pictures/}{images/}{./}} 

\usepackage{tabu}                      
\usepackage{booktabs}                  
\usepackage{lipsum}                    
\usepackage{mwe}                       

\usepackage{mathptmx}                  

\begin{document}


\firstsection{Introduction}

\maketitle

Network virtual reality (NVR) whiteboards provide an ability for remote collaboration, discussion, and immersive interaction. They have emerged as a promising alternative to physical and traditional electronic whiteboards ~\cite{ erofeeva2021impact,kumar2024cnn,petrykowski2019digital}. Unlike physical whiteboards, NVR whiteboards overcome spatial limitations, allowing users to interact in real-time regardless of location. Compared with traditional electronic whiteboards, which often provide limited and rigid interaction, NVR whiteboards use head-mounted displays (HMDs) with a large field of view (FOV) and support handwriting input via controllers and gesture recognition, resulting in an interaction experience that closely resembles real-world writing ~\cite{gronbaek2024blended,liu2025towards,merzhauser2025uninotesxr}. This design improves interaction comfort and realism while enabling real-time multi-user collaboration in a shared virtual space.

A key capability of NVR whiteboard systems is to support real-time immersive collaboration among remote users, as shown in \cref{fig:Fig1}. However, transmission and processing constraints inevitably introduce latency between users’ whiteboard inputs and their remote display, which may significantly degrade the user experience. Prior work has mainly compared the performance of VR-based and traditional electronic whiteboards in local environments ~\cite{erofeeva2021impact,jahangir2023identifying,minnekanti2024classesinvr,wang2021operation}. In contrast, the quality of experience (QoE) in NVR whiteboard collaboration, where latency critically affects interaction and coordination, remains insufficiently understood.

In whiteboard applications, users’ QoE can be evaluated from two aspects: pragmatic and hedonic ~\cite{akhtar2019multimedia,hameed2024holistic,moller2009taxonomy,wechsung2012measuring}. 
The pragmatic aspect emphasizes functional effectiveness and task support, encompassing dimensions such as interactivity, efficiency, and believability ~\cite{hameed2024holistic,moller2009taxonomy}. In contrast, the hedonic aspect highlights experiential qualities such as enjoyment, immersivity, explorability, and novelty ~\cite{hameed2024holistic, wang2022virtual}. Despite evidence that latency disrupts collaborative QoE, how it differentially affects these aspects, particularly whether its impact concentrates on specific sub-dimensions, remains unclear. This gap limits both the understanding of latency-driven QoE degradation and the design of NVR whiteboard systems resilient to latency.

However, studying the impact of latency on QoE in NVR whiteboards is challenging, as its effects depend not only on latency itself but also on contextual factors, particularly the collaboration mode and the platform form. In whiteboard applications, two common collaboration modes are sequential collaboration (SC) and free collaboration (FC). SC reflects structured workflows with alternating inputs, making interactions tightly coupled and thus more sensitive to latency-induced delays. In contrast, FC allows simultaneous contributions with looser coupling, offering greater flexibility but increasing the risk of missed, duplicated, or overlapping inputs under latency. These differences suggest that SC and FC exhibit distinct sensitivities to latency. Moreover, the presence or absence of digital avatars may further modulate users’ perception of latency and its impact on QoE.

This study investigates the impact of latency on the QoE of NVR whiteboards through a two-stage approach. First, QoE was classified into pragmatic and hedonic aspects, each comprising multiple sub-dimensions. Through controlled experiments, we analyze the sensitivity of these sub-dimensions to latency and identify three pragmatic dimensions, namely interactivity, efficiency, and believability, as the most affected, which are therefore adopted as the primary QoE evaluation indicators. Specifically, interactivity reflects the naturalness of interaction with the system and collaborators, efficiency captures the relationship between time and effort invested and task effectiveness, and believability denotes users’ trust in the system and their collaborators. Based on these QoE indicators, we designed two collaboration scenarios corresponding to the SC and FC modes. To examine platform-related effects, the scenarios were implemented on two VR whiteboard configurations: (1) a VR mode with visible user avatars and embodied actions (VR+), and (2) a VR mode without avatar representations, where only drawing strokes are visible (VR), enabling analysis of latency effects in the absence of avatar-related social cues. A traditional PC-based whiteboard was included as a baseline reference. This experimental design enabled a systematic evaluation of how latency influences QoE across different collaboration modes and platforms, providing insights into how latency shapes collaborative experiences and informing the optimization of NVR whiteboard systems with respect to real-world constraints. The main contributions of this paper are as follows:

\begin{itemize}
\item A self-controlled RCVR whiteboard system was developed, capable of precisely manipulating latency to reproduce real-world collaboration scenarios. While existing platforms provide useful functions, they often face limitations in latency control, task customization, parameter acquisition, and ratings scores collection for experimental purposes. The proposed system is designed as a flexible research tool and will be released as open source to facilitate future studies on latency-sensitive whiteboard interactions.

\item QoE was analyzed from both pragmatic and hedonic perspectives, with three pragmatic sub-dimensions, namely interactivity, efficiency, and believability, identified as particularly sensitive to latency. Building on this, the study further clarified their relative contributions to overall QoE under varying latency levels, thereby uncovering the mechanisms through which latency shapes collaborative whiteboard experience.

\item A systematic evaluation of latency effects was conducted by designing representative collaboration modes and implementing them across multiple platform types. This setup enabled a comprehensive comparison of QoE outcomes under different collaboration modes and platform settings.
\end{itemize}

The remainder of this paper is organized as follows. \cref{sec:Related Work and Motivations} reviews the related work and outlines the motivations of this study. \cref{sec:Design of Experiment} describes the experimental design, including the experimental platform, designed tasks, QoE metrics, participants, and procedures. \cref{sec:Analysis of Latency Effects on QoE} first presents a pilot study to identify QoE sub-dimensions sensitive to latency, and then analyzes the formal user study to examine the effects of latency on interactivity, efficiency, believability, and overall QoE across collaboration modes and whiteboard platforms. \cref{sec:Limitations} discusses the limitations of this study and outlines directions for future work. Finally, \cref{sec:Conclusions} concludes the paper.


\section{Related Work and Motivations}
\label{sec:Related Work and Motivations}

Whiteboards play an extremely important role in collaboration and communication, helping to bridge gaps in understanding through visual expression and shared interaction. With the development of immersive technologies, whiteboard systems have gradually evolved from traditional 2D forms to more immersive 3D space interaction tools. This section focuses on the relevant research on whiteboard systems, effect of latency on perception, and the motivations of this study.

\subsection{Whiteboard Systems}
\label{subsec:Whiteboard Systems}
The physical dry-erase whiteboard promotes local collaboration through natural handwriting and is an important auxiliary tool commonly used in social places such as work and study ~\cite{khairullah2024easy,mccorkle2020lightboard,rasmussen2013consider}. However, they do not support remote collaboration and their content is not inherently digitized for storage or sharing. Prior research on dry-erase whiteboards has investigated methods for digitizing whiteboard content
~\cite{varona2018post}.

With the development of communication technology and social media, electronic whiteboards are widely used in remote collaboration and meetings~\cite{gumienny2011tele,handley2023best,pikas2022digital,reguera2021using}. For instance, “Zoom Whiteboard” is a public whiteboard function in the remote meeting software “Zoom” ~\cite{reguera2021using}, while neXboard is a browser-based application offering a digital whiteboard for remote collaboration ~\cite{gumienny2011tele}. These electronic whiteboards are typically mouse-based, resulting in lower efficiency and discomfort during writing or drawing, which makes it challenging to replicate a natural handwriting experience. Another commonly used whiteboard device that supports remote collaboration is a digital tablet or a touch screen with a stylus. Rekimoto suggested using a 2D touch screen for public whiteboards for information sharing ~\cite{rekimoto1998multiple}. However, its size limitation is a problem. Extensive research has been conducted on electronic whiteboards. For example, ~\cite{ringe2015html5} developed a web application to increase collaboration without limiting it to any location, operating system platform, or device. The proposed application allows users to interact and share information through writings, images, chats, audio, and video. The authors in ~\cite{luo2016effect} investigated the perception of using different whiteboard interactive features and the effects of different whiteboard interactive uses on the learning attitudes. The results showed that using basic interactive features of whiteboard contributes to the cultivation of interest and willingness to learn.

In recent years, advancements in VR technology and hardware have brought significant attention to VR whiteboards ~\cite{abramczuk2023meet,chandana2024transforming,merzhauser2025uninotesxr,sakuraba20123d,tepljakov2021interactive}. Users can write and draw in virtual space through HMDs, breaking through the limitations of traditional physical space and objects to provide a more flexible and creative experience. For instance, Sakuraba et al. employed HMDs for collaborative 3D model design ~\cite{sakuraba20123d}. However, due to hardware limitations, VR whiteboards are often only single-machine operation and lack remote collaboration functions, which is one of the core requirements of whiteboard applications. As a result, research on remote collaborative VR whiteboards has become a prominent topic. Meta’s collaborative office application, Workrooms, includes a VR whiteboard feature that enables remote users to collaborate on a shared whiteboard ~\cite{abramczuk2023meet}. Recent studies have focused on developing and optimizing remote collaborative VR whiteboard systems. For example, ~\cite{hoff2023preparing} proposed a method for preparing for architectural software redesigns via hand-drawn sketching in VR whiteboard. In ~\cite{wang2022virtual}, the authors compared the collaborative efficiency of VR whiteboards and traditional electronic whiteboards. The results showed that VR whiteboards significantly improved collaboration efficiency compared to mouse-based electronic whiteboards and were comparable to traditional whiteboards. Nevertheless, commercial VR whiteboard systems such as Workrooms may pose certain limitations for experimental studies, particularly with respect to latency control, task customization, access to performance parameters, and the collection of subjective ratings scores.

\subsection{Effect of Latency on Perception}
\label{subsec:Effect of Latency on Perception}
Research in human-computer interaction (HCI) indicates that the threshold for acceptable latency varies significantly depending on the interaction context. Early and influential studies have systematically analyzed latency sensitivity across different user actions in networked games and virtual environments~\cite{claypool2006latency, steed2009networked}. For example, the tolerance is reported to be around 75 ms for full-body avatar movements ~\cite{waltemate2016impact}. In contrast, for tasks requiring character control in computer games, the acceptable range spans from 50 to 200ms ~\cite{claypool2006latency, jorg2012responsiveness,pantel2002impact}. For tasks where users press a button and then observe a visual stimulus, latencies below 50ms are typically imperceptible to users ~\cite{raaen2014can}. In addition, latency not only increases the perceived difficulty of tasks but also impacts adaptability. When tasks are predictable, users can adjust to latency more easily ~\cite{rohde2014predictability}. Latency affects the sense of presence in virtual environments, with higher latency leading to a diminished sense of presence ~\cite{meehan2003effect}. However, these findings are largely limited to HCI settings that focus on interactions between humans and systems, rather than on collaborative interactions between multiple users. Moreover, prior work has generally examined latency in relation to single QoE factors, while comprehensive analyses of how latency influences multiple QoE dimensions in collaborative contexts remain scarce.

To examine the impact of latency on user interaction in VR, Vlahovic et al.~\cite{vlahovic2019impact} investigated how network latency affects user experience in a first-person shooter VR multiplayer game. Their findings revealed that when round-trip time (RTT) exceeded approximately 100ms, the user experience for the selected game began to deteriorate. Similarly, Kojic et al.~\cite{kojic2019influence} explored the effects of varying latency levels on QoE in VR multiplayer sports games. They found that the perception of latency and QoE varied based on the users’ own latency levels. Participants could sense latency even when it only affected their opponents, and QoE scores significantly decreased only at higher latency levels. Kusonose~\cite{kusunose2010qoe} introduced a dual-user network air hockey game with haptic support, reporting that perceived interactivity and quality declined significantly as latency increased. Sithu et al.~\cite{sithu2015qoe} conducted a QoE evaluation of the operability and fairness of a two-player real-time balloon bursting game. Their results showed that operability was primarily dependent on network latency, with subjective ratings dropping below 3 when latency reached 300ms or higher. While these studies provide valuable insights into latency effects, they are centered on competitive scenarios whose characteristics do not fully align with collaborative work contexts. In applications such as whiteboard collaboration, where the emphasis is on teamwork and productivity, it is therefore essential to examine latency effects directly, as the nature of interaction may give rise to different sensitivities.

\subsection{Motivations}
\label{subsec:Motivations}
The motivation for this study can be summarized in three aspects:
\begin{itemize}
\item NVR whiteboards are particularly vulnerable to latency due to factors such as bandwidth limitations, concurrent users, and rendering complexity. Accurately assessing its impact is therefore essential, as it provides guidance for latency optimization. Once latency requirements are met sufficiently, service providers can further enhance performance by improving bandwidth efficiency, accommodating more users, or supporting more complex rendering.
\item Compared to traditional electronic whiteboards, NVR whiteboards offer greater freedom and immersion, which can alter how latency affects user experience. However, the specific QoE dimensions most sensitive to latency, particularly across pragmatic and hedonic aspects, remain unclear, making it necessary to investigate latency effects in NVR whiteboard scenarios.
\item In NVR whiteboards, collaborative modes (e.g. SC and FC) and platform conditions (e.g., with or without avatars) may influence user sensitivity to latency and overall QoE. However, these factors have not been systematically examined, highlighting the need for targeted studies to clarify how latency effects vary between collaborative modes and platform settings.
\end{itemize}

\section{Design of Experiment}
\label{sec:Design of Experiment}
To evaluate subjective QoE under different latency conditions, a series of experimental studies were conducted. We first developed an NVR whiteboard testing platform, which provides core collaboration functions and allows precise control of latency. Based on two representative whiteboard usage scenarios, namely SC for structured design workflows and FC for open discussion, corresponding experiments were designed. A pilot study was then carried out to identify the QoE sub-dimensions most sensitive to latency and most valued by users in collaborative whiteboard interaction. Based on these findings, a formal experiment was conducted to systematically collect subjective QoE ratings across different types of whiteboard, collaboration modes, and latency conditions.

\subsection{Experimental Platform}
\begin{figure}[tb]
 \centering 
 \includegraphics[width=\columnwidth]{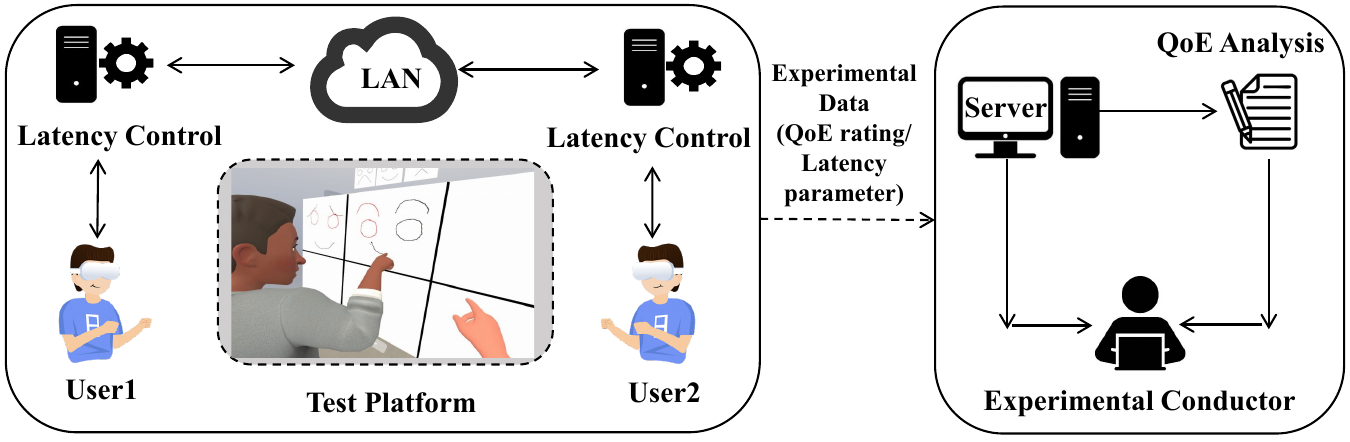}  
 \caption{Overview of the experimental setup and data processing workflow. Left panel: Collaborative whiteboard test platform with controlled network latency. Right panel: Centralized collection and processing of experimental data for subsequent QoE analysis.}
 \label{fig:Fig2}
\end{figure}

\begin{figure}[tb]
 \centering 
 \includegraphics[width=\columnwidth]{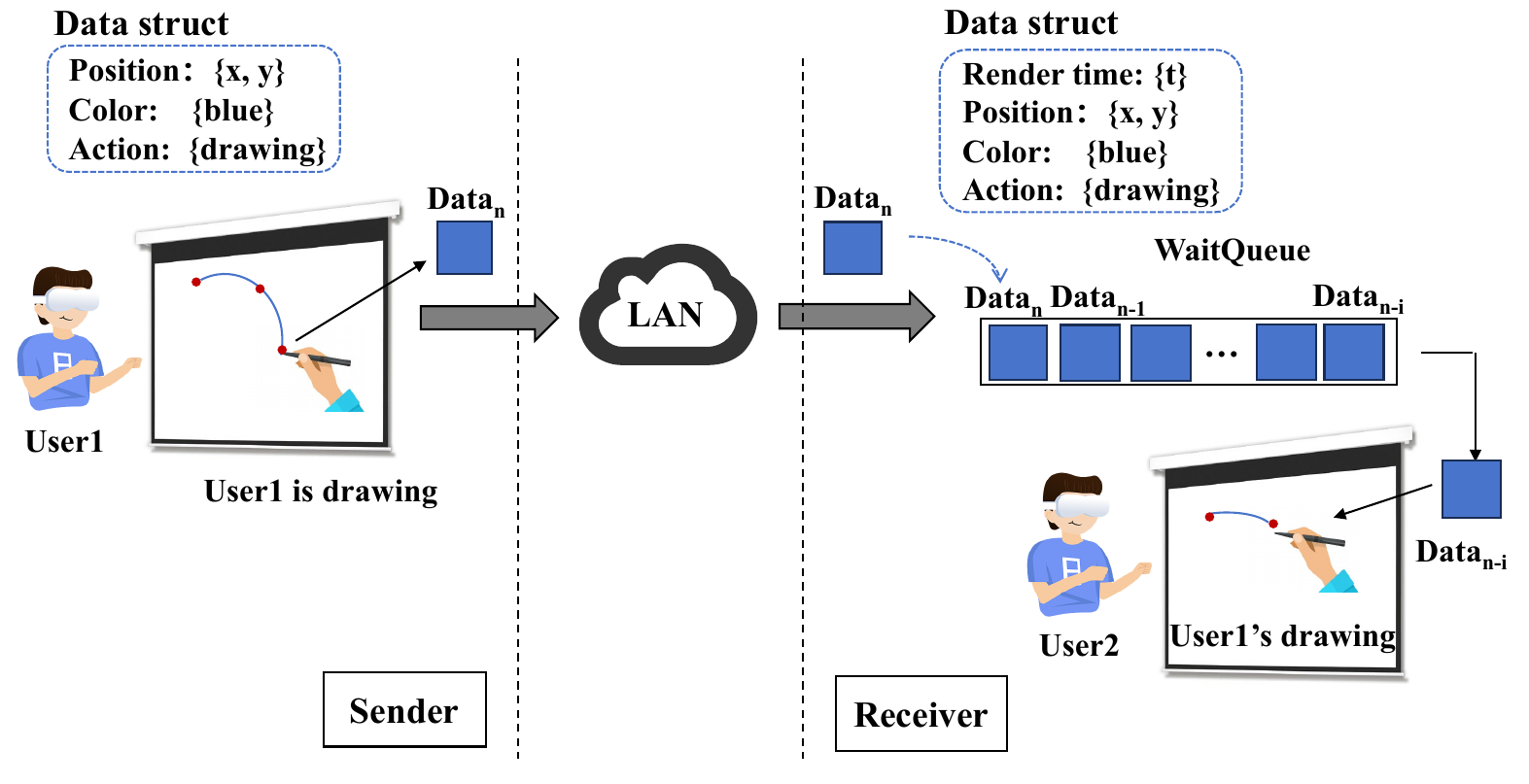}
 \caption{Adjustable E2E latency simulation using a time-based buffering mechanism.}
 \label{fig:Fig3}
\end{figure}

Considering the limitations of commercial VR whiteboard systems in latency control, task customization, parameter access, and subjective rating score collection, a latency-controllable experimental platform was developed in this study, accompanied by a centralized data collection (e.g., latency parameters and QoE ratings) and processing workflow for subsequent QoE analysis, as illustrated in \cref{fig:Fig2}. This platform was built based on four core elements: VR environment design, collaborative scenarios simulation, remote connection support, and adjustable simulated end-to-end (E2E) latency. Next, the functions and implementation of these four core elements will be described in detail.\\[1em]
\textbf{VR environment design:} 
The system was developed using the Unity engine and ran on the Meta Quest 2 with Meta Horizon. It provided a 3D virtual room with a shared whiteboard at the center. When a user entered the application and established a network connection with their partner, the system placed the two users on either side of the whiteboard to facilitate collaborative task execution. The user avatars were generated using the Meta Avatars SDK, with enhanced realism enabled by hand tracking technology and internal algorithms, as shown in \cref{fig:Fig1}. An earlier version of the Avatar SDK without lower-body representations was used. To ensure experimental consistency, avatar appearance was fixed, with only a binary gender selection (male/female) permitted. Predefined avatar models were then assigned deterministically based on the users’ initial positions in the virtual space, and no further avatar customization was allowed during the experiment. For comparison, the system allowed the avatar to be removed from the whiteboard environment. The detailed system configuration are listed \cref{tab:system_config}.\\[1em]
\textbf{Collaborative scenarios simulation:} The VR system was designed to simulate realistic whiteboard collaboration scenarios based on natural hand gestures. Users interacted with the whiteboard by using their right index finger to draw, erase, switch pen colors, or clear the surface, with all actions tracked and recorded in real time. To guide participants, a task template was displayed above the whiteboard, ensuring consistent task execution. Two typical collaboration scenarios were modeled to reflect real-world usage: in the SC scenario, participants alternated their inputs in a structured workflow, each building on their partner’s contribution. In the FC scenario, participants worked simultaneously, drawing and annotating in parallel to mimic open discussions. Together, these scenarios provided a controlled yet realistic simulation of collaborative whiteboard use, enabling systematic investigation of latency effects on user QoE. To maintain experimental control, vocal communication was not included in the current system, and the study focuses on visually dominated collaboration scenarios.\\[1em]
\textbf{Remote connection support:} The platform was designed to enable two users to collaboratively write and draw through a peer-to-peer connection, allowing them to observe each other’s avatar movements and whiteboard content. Each user’s avatar was driven by motion data captured by the HMD devices built-in tracking algorithms. The user’s real hand movements were captured through the HMD cameras and mapped onto the digital avatar. The system adopts a WebRTC-based peer-to-peer (P2P) communication model. Peer connections are established via a signaling server deployed within the local area network (LAN), after which stroke and avatar data are exchanged directly between devices through WebRTC DataChannels. The party’s movements were captured and transmitted through the LAN to drive the party’s digital avatar, realizing the collaboration of both parties in a shared virtual space. At the same time, whiteboard content data (coordinates, color, etc.) were transmitted alongside avatar data to ensure synchronized whiteboard interaction. The network connection and real-time data transmission were implemented using the WebRTC for Unity library~\cite{garcia2025rtc}.\\[1em]
\begin{table}[t]
\centering
\caption{System configuration of the experimental platform}
\label{tab:system_config}
\begin{tabular}{ll}
\hline
\textbf{Component} & \textbf{Specification} \\
\hline
VR Device & Meta Quest 2 \\
Operating System & Meta Horizon OS v74.1021 \\
Unity Engine &  Unity 2021.3.9f1 \\
Avatar SDK & Oculus Avatar SDK v20.1 \\
WebRTC Library & com.unity.webrtc 3.0.0-pre.4 \\
Networking Model & WebRTC-based peer-to-peer (P2P) \\
\hline
\end{tabular}
\end{table}
\textbf{Adjustable E2E latency simulation:} In order to achieve quantitative analysis of latency effects, the NVR whiteboard system implemented a precise E2E latency control mechanism. Operating within a LAN, the system’s baseline E2E latency was approximately 80 ms. On this basis, artificial latency was introduced through a time-based buffer mechanism, as illustrated in \cref{fig:Fig3}. When receiving real-time data (e.g., avatar motion or whiteboard strokes), the system recorded the arrival timestamp Tr and placed the data into a waiting queue (WaitQueue). The system then monitored this queue at a fixed rate of 60 frames per second. For each frame, the system checked the timestamp of the data at the front of the queue. When the following condition was satisfied, the data was dequeued and rendered in the virtual environment:
\begin{equation}
\vspace{0.3em}
Tc-Tr \geq Dt-Di
\end{equation}
where \textit {Tc} is the current system time, \textit{Tr} is the arrival time of the data, \textit{Di} is the inherent latency of the system. \textit{Dt} is the preset target latency. This mechanism ensured fine-grained control of total latency by accounting for the inherent delay, and was consistently applied to both interactive actions (e.g., drawing or gesture input). The fixed logical update rate of 60 Hz was used for processing buffered data during latency simulation. Although this rate differs from the native display refresh rate of the Meta Quest 2, the resulting temporal offset was minimal and did not noticeably affect interaction during the whiteboard tasks, which mainly involved slow and continuous user motions such as handwriting and arm movements.

For comparison, a traditional PC-based electronic whiteboard with controllable latency was also developed, as shown in \cref{fig:Fig1}. Built with Node.js, it supported various drawing functions via mouse input and allowed multi-user collaboration through socket.io-based data transmission, with a baseline system latency of approximately 27 ms. Its latency control mechanism was identical to that of the VR whiteboard system, including synchronized treatment of drawing events.

\subsection{Design of Collaboration}
To investigate the impact of latency on the QoE of NVR whiteboards under different conditions, we aimed to simulate realistic usage scenarios under controlled experimental settings. To this end, two collaboration scenarios were implemented to represent SC and FC, respectively. Participants were grouped in pairs and asked to jointly reproduce six predefined line-drawing templates within designated areas on a shared whiteboard (\cref{fig:Fig1}). Before drawing began, participants were allowed to use voice communication for coordination, such as signaling readiness to start. However, once the drawing process started, verbal discussion was not permitted, ensuring that collaboration relied solely on the whiteboard interaction. In the SC scenario, participants alternated strokes in a strict sequence, with each participant allowed to proceed only after their partner had completed the previous stroke, thereby enforcing tightly coupled collaboration. In the FC scenario, both participants drew simultaneously, freely selecting any unfinished stroke from any template without restrictions on order. The session was considered complete once all strokes across the templates had been finished.

\subsection{Participants}
A total of 36 participants (12 female and 24 male) were recruited for the study. The study consisted of two experimental phases: a pilot study and a formal user study. Eighteen participants (6 female and 12 male, aged 23-26, M = 23.83, SD = 0.69) took part in the pilot study to validate the experimental setup. During the formal phase, 24 participants (8 female and 16 male, aged 23-25, M = 23.58, SD = 0.49) were recruited for the main subjective evaluation, including 6 participants who had also participated in the pilot study. 

The participants were university students with normal or corrected-to-normal vision and hearing. This group was selected because university students are more likely to use VR whiteboards in educational and collaborative contexts, making them representative of a key user population. They also possess sufficient digital literacy to operate VR systems effectively, which minimizes confounding effects from lack of technical familiarity. Although this controlled participant selection may limit generalizability, the narrow age range and controlled sensory conditions reduce variability unrelated to the experimental manipulations, ensuring that the results primarily reflect the effects of latency and platform differences. Participants were divided into pairs to complete the collaborative tasks, consistent with the natural use of whiteboards.

The study involved a low-risk user experiment and did not include any invasive procedures or the collection of personally identifiable or sensitive data. All participants provided informed consent prior to participation. The experimental tasks were limited to VR-based or PC-based whiteboard collaboration conducted under controlled laboratory conditions, posing minimal risk to participants. Given the low-risk nature and scope of the study, the experimental protocol did not fall under formal institutional review board (IRB) review and was instead reviewed internally to ensure compliance with established ethical principles and institutional guidelines.

\subsection{Experimental Procedure}
The experiment followed a two-stage approach, consisting of a pilot study and a formal experiment. The pilot study had two main purposes: to familiarize participants with the testing procedure and to identify the QoE sub-dimensions most sensitive to latency and most cared by users in NVR whiteboard interaction. Building on these findings, the formal experiment was then conducted to systematically collect subjective QoE ratings under different latency conditions, across multiple whiteboard types and collaboration modes.

\subsubsection{Pilot experiment}
In the pilot study, QoE for NVR whiteboard applications was first classified into two aspects, namely pragmatic and hedonic. The pragmatic aspect included three sub-dimensions: interactivity, efficiency, and believability. The hedonic aspect included sub-dimensions such as enjoyment, immersivity, explorability, plausibility, and novelty. Participants evaluated their experience along these sub-dimensions under latency conditions. The definitions of each sub-dimension are described as follows:
\begin{itemize}
\item \textbf{Interactivity} refers to the degree of naturalness and fluency of users’ interaction with the whiteboard system and collaboration with their party.
\item \textbf{Efficiency}  reflects the degree of match between time invested in the task process and task completion.
\item \textbf{Believability}  captures the user’s sense of trust in the whiteboard system and their peers, as well as their confidence in the collaborative process.
\item \textbf{Enjoyment}  denotes the level of pleasure, relaxation, or ease users experience during the collaborative process.
\item \textbf{Immersivity}  refers to the extent to which users feel present or immersed in the whiteboard environment, as if they are physically part of the system.
\item \textbf{Explorability}  represents the freedom of movement of users in the whiteboard collaboration environment, as well as the sense of space and spaciousness.
\item \textbf{Plausibility} represents the degree to which the actions performed and the feedback received are consistent with the physical laws of the real world. (For example, whether the hand can directly pass through the whiteboard).
\item \textbf{Novelty} relates to the sense of newness or uniqueness experienced while using the whiteboard system.
\end{itemize}

During the pilot phase, we carried out an a priori power analysis, which determined that, assuming an effect size (f=0.25), a significance level ($\alpha$) of 0.05, a statistical power (1-$\beta$) of 0.95, and a 2×3 within-subjects design (Latency: no-latency vs. 1000 ms; Platform: VR+, VR, PC) evaluated across eight QoE sub-dimensions, a minimum of 18 participants was required. The participants were first introduced to the experimental objectives and procedures and received training on how to operate the different experimental platforms. Next, they experienced both no-latency and 1000 ms conditions on three types of collaborative whiteboards, namely VR+, VR and PC. Then, participants were asked to evaluate whether latency had a significant impact on each QoE sub-dimension (Yes/No) and whether users strongly cared about these sub-dimensions in the whiteboard collaboration (Yes/No).

\subsubsection{Formal experiment}
During the formal phase, we first conducted an a priori power analysis for the experimental design. Assuming a medium effect size (f=0.25), $\alpha$=0.05, a sample of 18 to 20 participants achieves 90-95\% power. As the collaborative tasks were performed in pairs, observations within each pair are not strictly independent. Therefore, the power analysis was conducted at the participant level as a practical approximation. Accordingly, 24 participants were employed in the subjective test, including six participants who had also attended the pilot study. Participants were randomly paired and placed in separate rooms to ensure that all communication occurred exclusively through the experimental platform. The sequence of the three experimental platforms was randomized for each pair. On each platform, participants engaged in two collaboration scenarios (SC and FC) presented in random order. Within each scenario, seven latency levels were tested: 100ms, 300ms, 600ms, 1000ms, 1500ms, 2000ms, and 2500ms. These latency levels were selected to span a wide range of conditions, including extreme cases, in order to capture the full degradation trend of user QoE under increasing latency. The order of these latency levels was randomized to minimize order effects. After completing one scenario at a given latency level, participants evaluated a set of QoE sub-dimensions using the 5-point Absolute Category Rating (ACR) method defined by ITU-T P.910~\cite{installations2023itu}. The selected sub-dimensions included interactivity, efficiency, believability, and overall QoE, which were informed by the pilot study. Each indicator was rated on a scale of “Perfect”, “Good”, “Fair”, “Poor”, and “Bad”, corresponding to scores from 5 to 1. Mean Opinion Scores (MOS)~\cite{installations2023itu} were then computed by averaging participants’ ratings for each condition. After completing all seven latency conditions on one platform, participants took a 15-minute break before proceeding to the next platform. This process was repeated until all platforms had been tested, at which point the experiment concluded.


\section{Analysis of Latency Effects on QoE}
\label{sec:Analysis of Latency Effects on QoE}

This section first reports a pilot study to identify QoE sub-dimensions most sensitive to latency. Based on these results, we further analyze the formal user study to examine latency effects on interactivity, efficiency, believability, and overall QoE across collaboration modes and whiteboard platforms, and further explore how these sub-dimensions relate to and shape overall QoE under varying latency conditions.

\subsection{Pilot Study Analysis}
Before examining latency effects, we conducted preliminary effect size analyses on the two sets of ratings, using Cohen’s h against a 0.5 baseline to assess the strength of participant agreement. For latency-affected ratings, large effects were observed in the pragmatic dimensions of efficiency (h = 1.57), believability (h = 0.85), and interactivity (h = 0.85), indicating consistent participant responses. Among the hedonic dimensions, enjoyment (h = 1.07) also showed a large effect, whereas immersivity (h = 0.39) was weaker, and the remaining dimensions were rated almost uniformly unaffected. For user-concern ratings, efficiency (h = 1.07) and believability (h = 0.85) again emerged as the strongest, with interactivity (h = 0.68) showing a moderate effect. By contrast, hedonic dimensions such as immersivity (h = -0.85) were considered important by some participants, while enjoyment and plausibility were weaker, and explorability and novelty showed little variance. Overall, pragmatic dimensions showed more consistent user ratings, while hedonic dimensions exhibited weaker and less reliable effects.

\begin{figure}[tb]
 \centering 
 \includegraphics[width=\columnwidth]{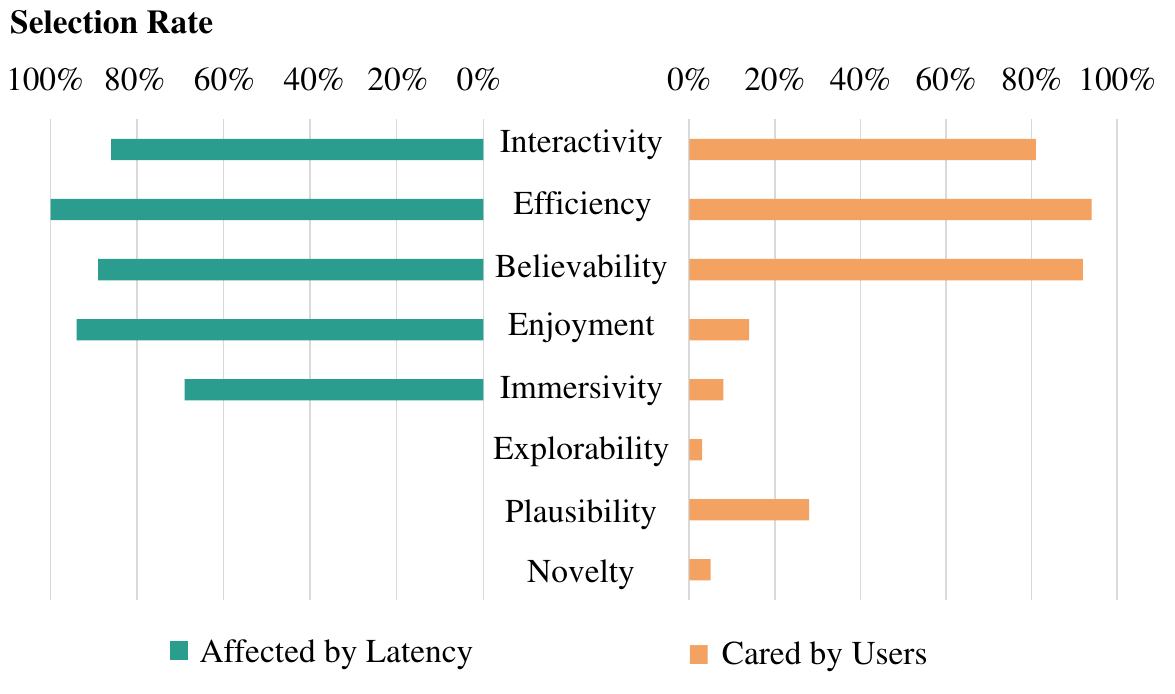}
 \vspace{-12pt}
 \caption{Rating results of pilot test.}
 \label{fig:Fig4}
\end{figure}

The results show that users prioritize certain QoE dimensions when these are closely related to the main tasks performed on collaborative whiteboards. As illustrated in \cref{fig:Fig4}, efficiency was rated by all participants as being affected by latency (100\%), and nearly all participants also considered it important (94\%). Similarly, believability was judged both highly affected (89\%) and strongly valued (91\%), highlighting its role in maintaining trust and consistency in collaboration. Interactivity also demonstrated a high proportion of impact (86\%) and concern (81\%), indicating that disruptions to responsiveness and synchronization directly influence users’ experience. In contrast, enjoyment presented an interesting discrepancy: although most participants perceived it to be affected by latency (94\%), only a minority (17\%) regarded it as important, suggesting that while latency may reduce fun, they are not considered critical in task-oriented collaboration. A similar pattern was observed for immersivity, where 67\% reported latency effects but only 9\% expressed concern, implying that immersivity are less prioritized than effectiveness. For the remaining hedonic dimensions, explorability, plausibility, and novelty, almost no latency effects were reported (0\%) and few participants expressed concern (6$\sim$28\%). These results indicate that such sub-dimensions are relatively peripheral in whiteboard-based collaboration.

Overall, the findings suggest that users focus primarily on pragmatic dimensions, including efficiency, believability, and interactivity, as these dimensions are not only highly sensitive to latency but also central to effective task completion and collaborative coherence. In contrast, hedonic dimensions such as enjoyment and immersivity, while occasionally perceived as affected, play a more peripheral role in shaping user priorities. This indicates that in latency-sensitive environments such as NVR whiteboards, users tend to prioritize dimensions that directly support functional performance and reliable coordination over those that mainly enhance experiential qualities. These insights highlight the need for system designers to optimize latency handling strategies around pragmatic dimensions to safeguard core collaborative effectiveness, while recognizing that hedonic aspects, although secondary, may still contribute to overall user satisfaction once fundamental functional demands are met.

\begin{figure*}[htbp]
  \centering
  \includegraphics[width=\linewidth]{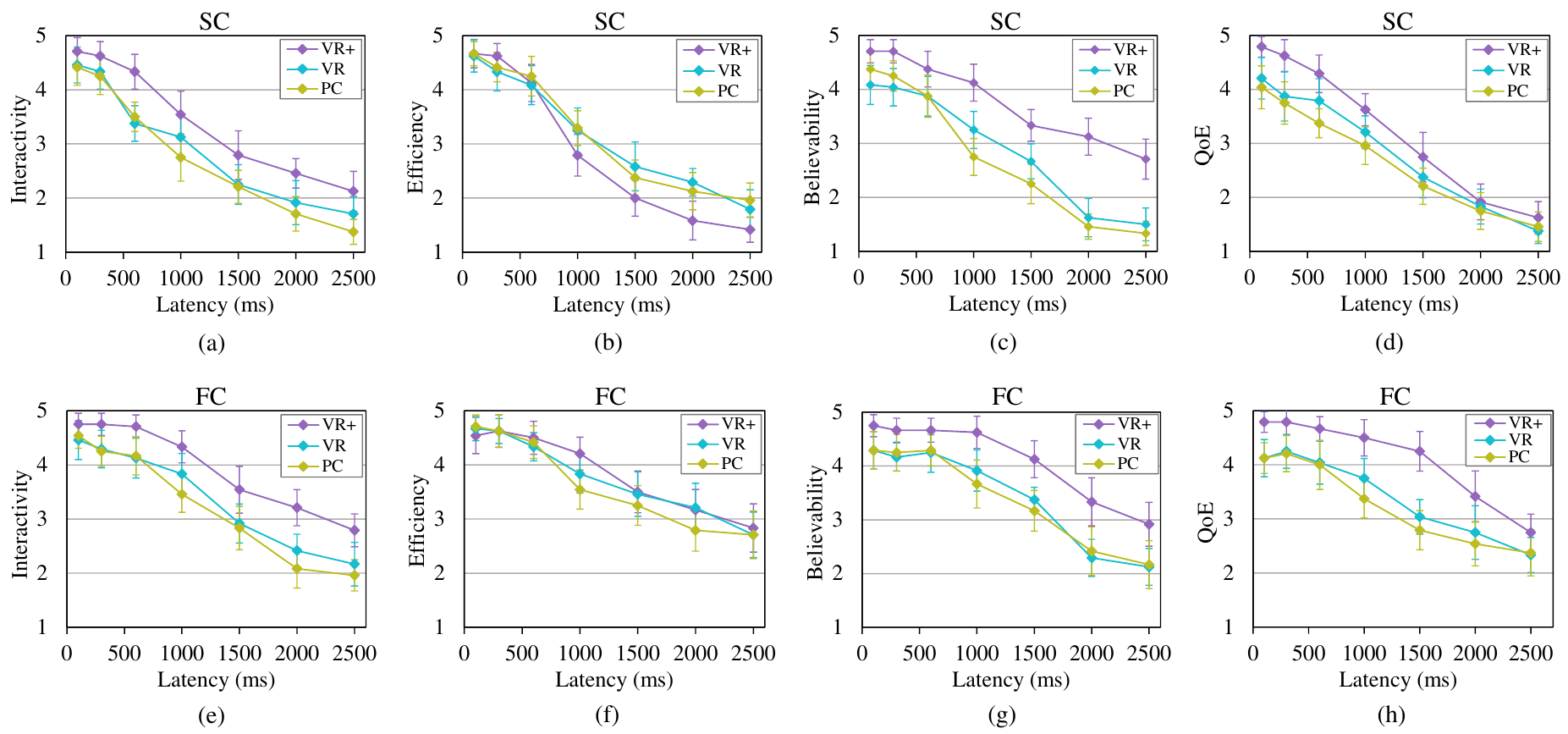}
  \caption{Comparison of the impact of latency on QoE across different whiteboard platforms. (a) Interactivity of SC, (b) Efficiency of SC , (c) Believability of SC, (d) QoE of SC, (e) Interactivity of FC, (f) Efficiency of FC, (g) Believability of FC, (h) QoE of FC (95\% confidence interval).}
  \label{fig:Fig5}
\end{figure*}

\begin{figure*}[htbp]
  \centering
  \includegraphics[width=\linewidth]{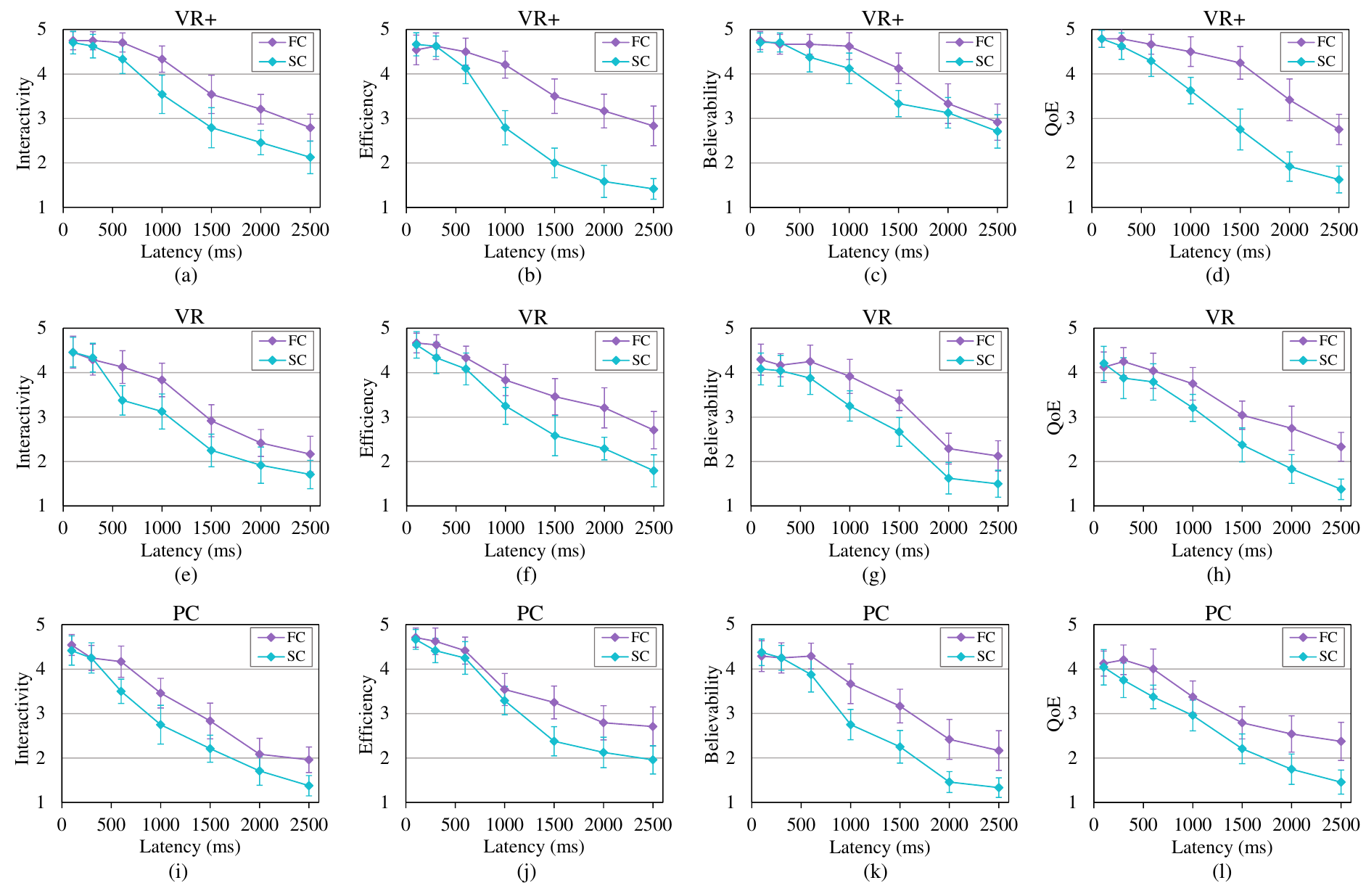}
  \caption{Comparison of impact of latency on QoE in different collaborative scenarios. (a) Interactivity on VR+ platform (b) Efficiency on VR+ platform, (c) Believability on VR+ platform, (d) QoE on VR+ platform, (e) Interactivity on VR platform (f) Efficiency on VR platform, (g) Believability on VR platform, (h) QoE on VR platform, (i) Interactivity on PC platform, (j) Efficiency on PC platform, (k) Believability on PC platform, (l) QoE on PC platform (95\% confidence interval).}
  \label{fig:Fig6}
\end{figure*}


\subsection{Impact Analysis on Interactivity}  %
The Shapiro-Wilk test~\cite{hanusz2016shapiro} was first applied to assess the normality of interactivity scores across collaboration modes (SC and FC) and platforms (VR+, VR, and PC). The results indicated that the data followed a normal distribution (all p \textgreater 0.05). Therefore, repeated-measures ANOVA~\cite{keselman2001analysis} was performed to further examine differences under two conditions: (1) within the same whiteboard platform across different collaboration modes, and (2) within the same mode across different whiteboard platforms.

As shown in \cref{fig:Fig5}a and \cref{fig:Fig5}e, interactivity declined with increasing latency across all platforms for both SC and FC scenarios. Nevertheless, the VR+ platform consistently demonstrates significantly higher interactivity compared to the other two platforms (F=33.99, p \textless 0.01, $\eta^{2}$=0.85). This advantage can be attributed to the use of digital avatars, which allow users to perceive their partner’s movements, postures, and spatial positions in the virtual environment. Such embodiment more closely simulates real-world whiteboard collaboration, thereby enhancing the sense of interactivity. By contrast, when latencies exceeded 300 ms, the VR platform yielded slightly higher interactivity scores compared to the PC platform, but this difference did not reach statistical significance (F=3.20, p=0.15, $\eta^{2}$=0.44), indicating that the observed advantage was not robust under higher-latency conditions. This finding indicates that a VR environment alone may be insufficient to improve collaborative interactivity. Instead, the inclusion of avatars emerges as the key element contributing to enhanced interactivity.

\begin{table*}[htbp]
  \centering
  \caption{Summary of QoE Findings under Different Latency Conditions}
  \label{tab:qoe_summary}
  \begin{tabular}{p{2.8cm} p{4.5cm} p{4.2cm} p{4.5cm}}
    \toprule
    \textbf{Condition} & \textbf{Sequential Collaboration (SC)} & \textbf{Free Collaboration (FC)} & \textbf{Implications} \\
    \midrule
    Low latency ($\leq$600 ms) &
    QoE remains high across all platforms, with VR+ consistently showing the highest ratings. &
    QoE is stable across platforms, but VR+ shows a distinct advantage. &
    Collaboration is reliable in this range, and avatar support can further enhance user experience. \\ [0.9 cm]
    
    Moderate latency 
    (600--2000 ms) &
    QoE deteriorates quickly, and collaboration coherence breaks down. &
    FC remains comparatively resilient and still usable.. &
    600 ms seems to be an important threshold. SC should be kept below this limit. \\[0.9 cm]
    
    High latency (\textgreater 2000 ms) &
    QoE collapses across platforms, and differences become negligible. &
    QoE is still poor but slightly higher than SC. &
    Collaboration becomes impractical, and such extreme delays must be avoided. \\[0.9 cm]
    
    Platform effects &
    VR+ outperforms VR and PC under low-to-moderate latency, but loses its advantage at extreme delays. &
    VR+ performs best under low-to-moderate latency, VR and PC show broadly similar patterns. &
    Avatars improve trust and interactivity, but their benefit is most evident when latency is kept within a manageable range. \\[0.9 cm]
    \bottomrule
  \end{tabular}
\end{table*}

From a collaborative mode perspective, the impact of latency on interactivity differs between scenarios across all platforms, as shown in \cref{fig:Fig6}a, \cref{fig:Fig6}e, and \cref{fig:Fig6}i. When latency is low ( \textless 300ms), no significant difference in interactivity is observed between FC and SC scenarios (p \textgreater 0.05). However, as latency increases ( \textgreater 300ms), users report higher interactivity in the FC scenario compared to the SC scenario (F=77.58, p \textless 0.01, $\eta^{2}$=0.95), and this difference becomes more pronounced with greater latency. This trend is consistent across all three platforms and reflects the underlying interaction dynamics: SC relies on tightly coupled turn-taking, where each action depends on the partner’s immediate input, making it highly vulnerable to disruptions once latency rises. By contrast, FC adopts a looser, parallel interaction structure, allowing participants to adjust asynchronously and tolerate moderate latencies. Overall, these findings suggest that collaboration mode determines sensitivity to latency, with 300 ms emerging as a important threshold for divergence.

\subsection{Impact Analysis on Efficiency}
Efficiency reflects users’ perceptions of task performance, particularly their ability to accomplish tasks effectively under varying levels of latency. As shown in \cref{fig:Fig5}b, efficiency ratings in the SC scenario decreased steadily across all three platforms as latency increased. When latency was low ($\leq$ 300 ms), the impact was minimal, with scores remaining around 4.3$\sim$4.7 and no significant platform differences (F=1.12, p\textgreater 0.05, $\eta^{2}$=0.09). By 1000 ms, however, VR+ dropped to 2.79, significantly lower than VR (3.25) and PC (3.29), and this divergence became more pronounced at 1500-2000 ms, when VR+ efficiency fell further (1.6$\sim$2.0) while VR and PC remained moderately higher (2.1$\sim$2.6). A repeated-measures ANOVA confirmed a significant main effect of platform under higher latency conditions (F=6.84, p \textless 0.01, $\eta^{2}$=0.36), indicating that VR+ was disproportionately affected. A possible explanation is that in VR+, the additional synchronization of avatars and embodied cues amplifies the perceptual salience of delay, making inefficiency more apparent when latency grows.

By contrast, in the FC scenario, as shown in \cref{fig:Fig5}f, the scores decrease with increasing latency, but the decline was more gradual and less dependent on platform. When latency is lower than 600ms, its impact on efficiency is also minimal, with efficiency scores declining at a relatively slow rate. Even at 1000 ms, VR+ (4.21), VR (3.83), and PC (3.54) maintained relatively high levels compared to SC. Repeated-measures ANOVA showed no significant differences between platforms across all latency levels (F=0.94, p \textgreater 0.05, $\eta^{2}$=0.07). This suggests that the loosely coupled, parallel nature of FC, where participants mainly focus on their own strokes with limited reliance on immediate partner feedback, buffered against latency, making efficiency more stable across platforms.

Direct comparisons between SC and FC (\cref{fig:Fig6}b, \cref{fig:Fig6}f, and \cref{fig:Fig6}j) highlight collaboration-related sensitivity. When latency is low ($\leq$ 600 ms), SC and FC efficiencies were nearly identical (p \textgreater 0.05). Beyond 600 ms, SC ratings declined more steeply, producing a significant interaction between collaboration modes and latency (F=12.47, p \textless 0.05, $\eta^{2}$=0.51). By 2000 to 2500 ms, SC dropped to 1.4$\sim$2.0, while FC remained at 2.7$\sim$3.2, showing that FC was consistently more resilient. The efficiency gap was largest on VR+, moderate on VR, and smallest on PC, reflecting the degree of coupling and synchronization cues available in each platform.

Overall, efficiency is highly sensitive to latency, but its decline is shaped by both collaboration mode and platform. SC, with tightly coupled turn-taking, is more fragile under latency, while FC maintains higher efficiency even at moderate latencies. Among platforms, VR+ suffered sharper declines under high latency, likely due to the added salience of delayed avatar synchronization. These findings suggest that to preserve efficiency, latency should be maintained below 600 ms, and interface designs should aim to reduce the impact of delayed feedback in avatar-supported VR.

\subsection{Impact Analysis on Believability}
As shown in \cref{fig:Fig5}c, believability in the SC scenario decreased steadily across platforms as latency increased. When latency was low ($\leq$ 300 ms), believability remained high (4.0$\sim$4.7) and showed little change across platforms. However, once latency exceeded 600 ms, the decline accelerated and platform differences became more evident. By 1000 ms, VR+ still maintained relatively high believability (4.13), while VR dropped to 3.25 and PC to 2.75. At 1500 to 2000 ms, VR and PC declined sharply (1.5$\sim$2.3), whereas VR+ retained moderately higher scores (3.1$\sim$3.3). A repeated-measures ANOVA confirmed a significant main effect of platform at higher latencies (F=8.27, p \textless 0.01, $\eta^{2}$=0.41), indicating that VR+ preserved believability better than VR and PC. A likely explanation is that avatars in VR+ provide additional social cues, such as visible partner presence and actions, which help sustain trust and credibility of collaboration even under delayed responses.

In FC scenarios, as shown in \cref{fig:Fig5}g, believability also decreased as latency increased, but VR+ consistently maintained higher values compared to VR and PC. When latency was below 600 ms, all platforms performed well (\textgreater 4.1) and scores remained stable. Beyond 1000 ms, however, differences became more apparent: VR+ still retained high believability (4.6), while VR and PC declined to 3.7 and 3.9, respectively. At 2000 to 2500 ms, the gap widened further, with VR+ remaining above 2.9, whereas VR and PC dropped to 2.1$\sim$2.4. Statistical analysis confirmed a significant main effect of platform at higher latencies (F=7.65, p \textless 0.01, $\eta^{2}$=0.39), showing that VR+ better sustained trust and engagement under latency. This suggests that in FC, where precise timing is less critical, avatars in VR+ further enhance believability by making partner engagement more visible.

\cref{fig:Fig6}c, \cref{fig:Fig6}g, and \cref{fig:Fig6}k compare the changes in perceived believability with latency in different collaboration modes on each platform. It indicates that believability in SC is more sensitive to latency than in FC. When latency is $\leq$600 ms, believability scores in the two scenarios were comparable. As latency increased, however, SC ratings declined more steeply, resulting in a significant mode-latency interaction (F=10.92, p \textless 0.01, $\eta^{2}$=0.48). The believability gap between SC and FC was most evident on PC and VR, where SC dropped more sharply under higher latency, while FC remained relatively stable. On VR+, believability scores were consistently higher across both scenarios, resulting in a smaller gap. 

Overall, believability decreases with increasing latency, but the rate of decline depends on both interactive modes and platforms. SC is more fragile due to its reliance on synchronous partner responses, while FC is more resilient thanks to its looser structure. Across both scenarios, VR+ consistently preserved higher believability, suggesting that avatar-based embodiment offers important social cues that help sustain believability even under delayed conditions.

\begin{table*}[htbp]
  \centering
  \caption{Dominant QoE Sub-Dimensions under Different Latency Levels and Collaboration Modes}
  \label{tab:qoe_subdimensions}
  \begin{tabular}{p{3cm} p{4.3cm} p{4.2cm} p{4.6cm}}
    \toprule
    \textbf{Latency range} & \textbf{Sequential Collaboration (SC)} & \textbf{Free Collaboration (FC)} & \textbf{Interpretation} \\
    \midrule
    Low latency ($<$300 ms) &
    All sub-dimensions remain high and QoE is stable. &
    All sub-dimensions remain high and QoE is stable. &
    Users perceive minimal disruption, and the impact of latency is largely negligible. \\[0.9 cm]
    
    Moderate latency (600--2000 ms) &
    \textbf{Interactivity dominates}: QoE is most closely aligned with interactivity. &
    \textbf{Believability dominates}: QoE is most closely aligned with trust and partner engagement. &
    SC users are highly sensitive to interactive flow, whereas FC users focus more on whether the partner appears attentive.\\[1.2 cm]
    
    High latency ($>$2000 ms) &
    \textbf{Efficiency dominates}: task completion ability becomes decisive as delays severely hinder progress. &
    Believability remains the most influential dimension, although overall QoE declines. &
    When responsiveness collapses, SC users judge QoE by efficiency, while FC users still weigh credibility of partner engagement. \\[0.9 cm]
    \bottomrule
  \end{tabular}
\end{table*}

\subsection{Impact Analysis on Overall QoE}
In this section, we first analyze the impact of latency on overall QoE under different collaboration modes and platform types. On this basis, we further analyze the relationship between each sub-dimension and overall QoE, and clarify the process of how sub-dimensions affect QoE under different latency.

\subsubsection{QoE under different modes and platforms}
As shown in \cref{fig:Fig5}(d), QoE ratings in the SC scenario decreased sharply as latency increased across all three platforms. At low latency ($\leq$300 ms), scores remained relatively high (3.8$\sim$4.8), with VR+ consistently outperforming VR and PC. At 600 ms, VR+ still held a score above 4.2, while VR dropped to 3.8 and PC to 3.4. Once latency reached 1000 ms, QoE declined more steeply, with VR+ at 3.63, VR at 3.21, and PC at 2.96. At higher latencies (2000$\sim$2500 ms), QoE scores dropped to very low levels (1.4$\sim$1.9) across all platforms. Although VR+ remained slightly higher than VR and PC, the advantage was no longer meaningful, as QoE had degraded to a level where platform differences became negligible.

In contrast, on FC scenarios, as shown in \cref{fig:Fig5}(h), QoE scores decreased with increasing latency, but the decline was more gradual compared to SC. At $\geq$600 ms, QoE remained consistently high (\textgreater4.0) across all platforms, with VR+ achieving the highest score (4.7$\sim$4.8). As latency increased to 1500$\sim$2000 ms, VR+ still maintained relatively higher score (3.4$\sim$4.3), while VR and PC dropped more steeply to around 2.5$\sim$3.0. By 2500 ms, all platforms had deteriorated, with VR+ at 2.75 and VR and PC around 2.3.

Comparisons across collaboration modes (\cref{fig:Fig6}d, \cref{fig:Fig6}h and \cref{fig:Fig6}l) show that FC was more resilient to latency than SC. At $\leq$300 ms, SC and FC showed similar QoE levels across platforms. However, as latency increased, SC scores declined faster, leading to a significant mode-latency interaction (F=11.8, p \textless 0.01, $\eta^{2}$=0.5). This indicates that the looser, parallel structure of FC allowed users to sustain better QoE under latency conditions, whereas the tightly coupled nature of SC made it more vulnerable.

Overall, a summary of QoE finding under different latancy conditions is list in \cref{tab:qoe_summary}. QoE decreases significantly as latency grows, but the collaboration mode and platform type both modulate this effect. SC is more fragile, showing a steep drop beyond 600 ms, while FC is more resilient. VR+ consistently achieved higher QoE than VR and PC, especially under higher latencies, likely because avatars provided additional social and contextual cues that preserved users’ engagement and trust despite slower responses. These findings suggest that maintaining E2E latency below 600 ms is critical for SC, while FC scenarios can tolerate somewhat higher latencies, particularly on avatar-supported VR+ platforms.

\subsubsection{Relationship between sub-dimensions and QoE}
To further explore how individual QoE sub-dimensions relate to and shape overall QoE under varying latency conditions, we focus on the VR+ whiteboard condition. Correlation analysis reveals strong associations between overall QoE and its sub-dimensions across collaboration modes (r \textgreater 0.95).

\begin{figure}[tb]
  \centering
  \includegraphics[width=0.9\columnwidth]{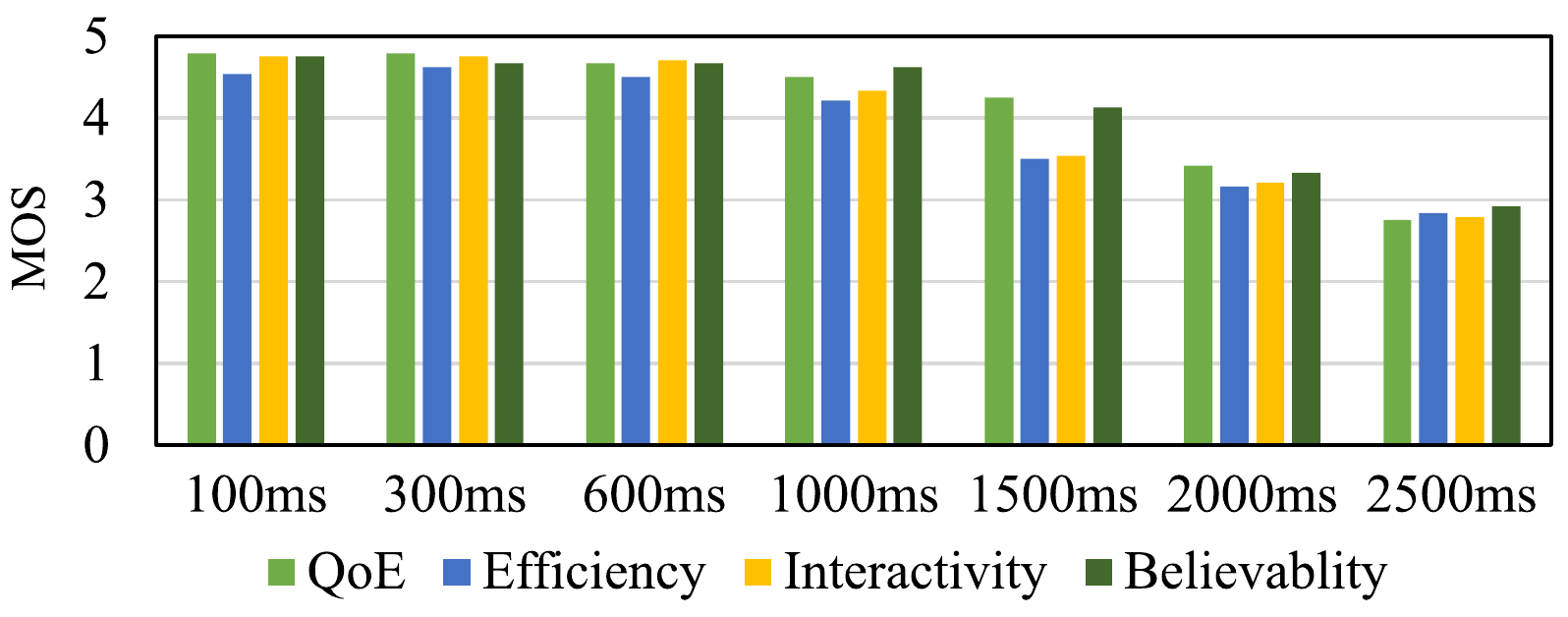}
  \vspace{0.1em}
  
  {\small (a) }
   \vspace{0.3em}

  \includegraphics[width=0.9\columnwidth]{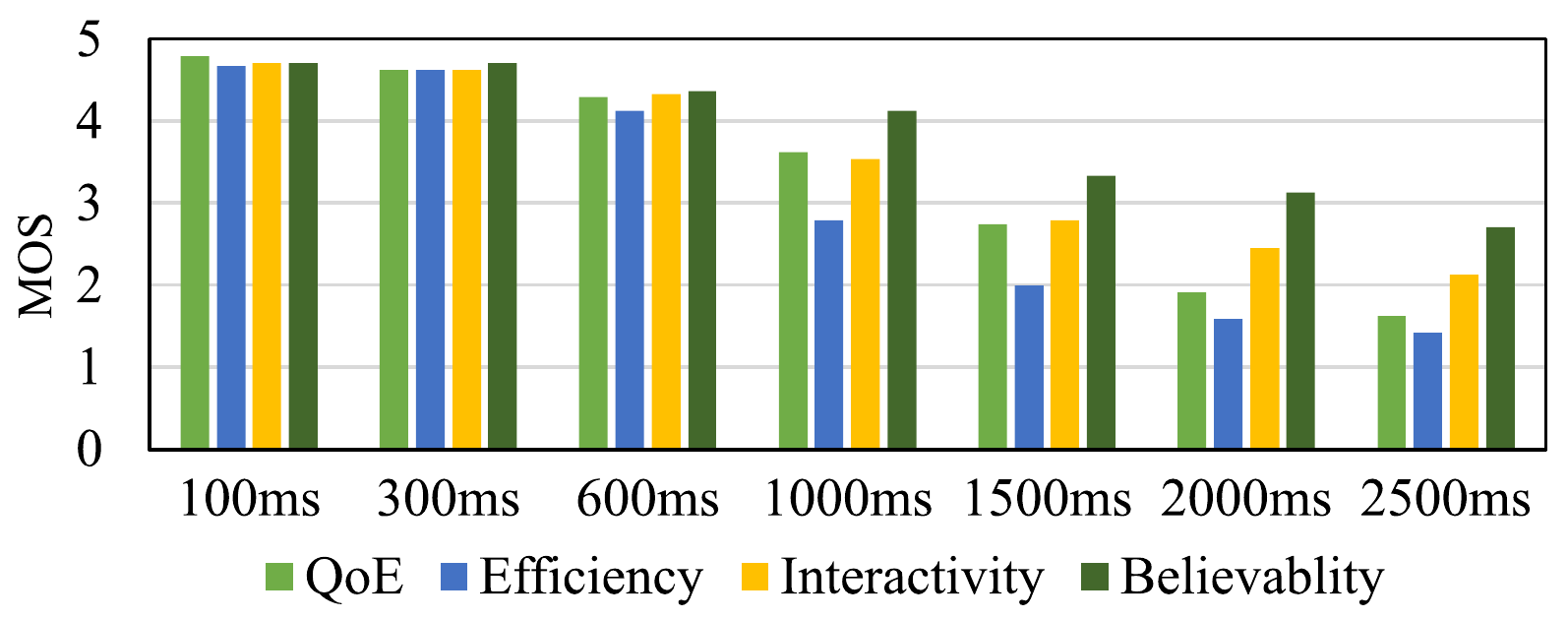}
  \vspace{0.1em}
  
  {\small (b) }
  \caption{Comparison between sub-dimensions and QoE for VR+: (a) SC and (b) FC.}
  \label{fig:Fig7}
\end{figure}

As shown in \cref{fig:Fig7}a, during the SC scenarios, when latency was low (\textless300 ms), all sub-dimensions remained high, resulting in consistently high QoE. As latency increased, however, QoE dropped significantly, primarily driven by declines in efficiency and interactivity. Notably, between 600 ms and 2000 ms, QoE was most strongly aligned with interactivity (T-test, p \textgreater 0.05), suggesting that under moderate latency, users were particularly sensitive to disruptions in real-time responsiveness and turn-taking. Once latency exceeded 2000 ms, efficiency became the dominant predictor of QoE (T-test, p \textgreater 0.05), indicating that when performance degradation was severe, users prioritized whether tasks could still be completed within reasonable effort and time. This reflects the fact that SC scenarios, being tightly coupled, demand both responsiveness and productivity, but the balance shifts from interactivity to efficiency as latency becomes intolerable.

In FC scenarios, as shown in \cref{fig:Fig7}b, QoE showed the strongest alignment with believability (T-test, p \textgreater 0.05). While efficiency and interactivity remained important contributing factors, the looser and more parallel structure of FC interactions meant that latency-induced disruptions generally stayed within users’ tolerance range. Instead, believability emerged as the decisive factor shaping overall QoE. Even when responses were delayed, avatars allowed participants to perceive their partner’s involvement, sustaining trust in the collaborative process. Thus, in FC scenarios, changes in believability had a disproportionately greater impact on QoE compared to other dimensions.

These findings highlight that different sub-dimensions dominate QoE depending on both latency level and collaboration mode. A summary of the dominant QoE sub-dimensions is list in \cref{tab:qoe_subdimensions}. In SC scenarios, interactivity is most critical under moderate latency, whereas efficiency dominates under severe latency. In FC scenarios, believability emerges as the key factor, as trust in partner engagement outweighs moderate efficiency losses. This differentiation not only clarifies the mechanisms by which latency affects QoE, but also provides actionable insights: systems supporting SC should focus on minimizing interaction lag below 600 ms, while those supporting FC should ensure mechanisms that strengthen partner visibility and trust, such as the use of avatars, even under constant network conditions.

 
\section{Limitations and Future Work}
\label{sec:Limitations}
This section summarizes the key limitations of the present study and highlights directions for future work aimed at extending the scope of the findings.

\begin{itemize}
\item\textbf{Participant Diversity}: The participant sample consisted of university students within a relatively narrow age range. While this controlled selection helps reduce variability unrelated to the experimental manipulations, it may limit the external validity of the findings for broader and more heterogeneous user populations. 
\item\textbf{Interaction Modalities}: Vocal communication was not included in the current study, and its potential interaction with latency effects remains to be explored.
\item\textbf{Network Conditions}: The experiments were conducted under controlled LAN conditions. Although this enables precise manipulation of E2E latency, real-world network environments may introduce additional factors, such as jitter and packet loss, which could further affect the user experience and therefore merit consideration in practical deployment scenarios.
\item\textbf{Evaluation Metrics}: The evaluation in this study primarily relies on subjective QoE ratings. While subjective assessments are essential for capturing user perception, the lack of objective performance metrics, such as task completion time, error rates, or interaction overlaps, limits a more systematic characterization of latency effects. 

\end{itemize}

Future work will extend the study to more diverse user groups, interaction modalities, and network conditions to further validate and generalize the findings. In addition, future research will focus on incorporating objective performance metrics and developing objective evaluation models to more systematically assess the effects of latency on QoE in NVR whiteboard collaboration.

\section{Conclusions}
\label{sec:Conclusions}
This study investigated how users’ QoE in NVR whiteboards varies across different latency conditions, while uncovering the key factors driving these changes. The findings indicate that, compared with hedonic aspects, users in NVR whiteboard collaboration prioritize pragmatic dimensions such as interactivity, efficiency, and believability. Latency impairs overall QoE primarily by disrupting these dimensions. At low latency, both SC and FC scenarios offered good user experiences, and the introduction of avatars further enhanced collaboration by strengthening interactivity and trust. At moderate latency, QoE in SC was most closely associated with interactivity, while in FC believability played a dominant role. In this stage, avatars remained beneficial, as they supported both responsiveness and credibility. At high latency, however, delayed avatar responses tended to amplify users’ negative perceptions of efficiency. Therefore, in highly interactive scenarios with severe latencies, reducing or omitting avatar use may therefore be preferable when efficiency is the primary concern. By adopting a user-perceived QoE perspective, this work provides practical guidance for optimizing NVR whiteboard systems under different network conditions. 

\acknowledgments{%
	This work was supported in part by the National Natural Science Foundation of China (62171353). %
}

\bibliographystyle{abbrv-doi-hyperref}

\bibliography{template}

\end{document}